

\titlepage \date{October,1992}
\voffset=24pt
\hfill{TIFR-TH-92/62}\break
\title{MATRIX MODELS AND BLACK HOLES}
\author{Sumit R. Das\foot
{e-mail: das@tifrvax.bitnet}}
\address{ Tata Institute of Fundamental Research, Homi Bhabha Road, Bombay
400 005, India}
\abstract{We show that an integral transform of the fluctuations of the
collective field of
the $d=1$ matrix model satisfy the same linearized equation as that of
the massless "tachyon" in the black hole background of the two dimensional
critical string. This suggests that the $d=1$ matrix model may provide
a non-perturbative description of black holes in two dimensional
string theory.}

\vfill
\endpage

\def\pax{\partial_x}
\def\half{{1 \over 2}}
\def\pat{\partial_t}
\def\tphi{{\tilde \phi}}
\def\pap{\partial_p}
\def\smu{{\sqrt{\mu}}}

\def\twa{{2 \over \alpha '}}
\def\pau{\partial_u}
\def\pav{\partial_v}
\def\tPi{{\tilde \Pi_\phi}}
\def\Win{W_{1+\infty}}
\def\win{w_{1+\infty}}
\def\stwo{{\sqrt{2}}}
\def\Pip{\Pi_\phi}
\def\onbeta{{1 \over \beta}}
\def\tom{T_\omega}
\def\unu{U_\nu}
\def\rega{{\rm Region~I}}
\def\regb{{\rm Region~II}}

\Ref\DONEG{D. Gross and N. Milkovic, Phys. Lett. 238B (1990) 217; E.
Brezin, V. Kazakov and Al. B. Zamolodchikov, Nucl. Phys. B338 (1990) 673;
G. Parisi, Phys. Lett. 238B (1990) 209; P. Ginsparg and J. Zinn-Justin,
Phys. Lett. 240B (1990) 333.}

\Ref\JEVSAK{A. Jevicki and B. Sakita, Nucl. Phys. B165 (1980) 511.}

\Ref\DJEV{S.R. Das and A. Jevicki, Mod. Phys. Lett. A5 (1990) 1639.}

\Ref\SW{A. Sengupta and S.R. Wadia, Int. J. Mod. Phys. A6 (1991) 1961; D.
Gross and I. Klebanov, Nucl. Phys. B352 (1990) 671.}

\Ref\MSW{G. Mandal, A. Sengupta and S.R. Wadia, Mod. Phys. Lett. A6
(1991) 1465;  K. Demeterfi, A. Jevicki and J.P. Rodrigues, Nucl. Phys. B362
(1991) 173 and Nucl. Phys. B365 (1991) 199; J. Polchinski, Nucl. Phys. B362
(1991) 125; G. Moore, Nucl. Phys. B368 (1992) 557; G. Moore, M.R. Plesser
and S. Ramgoolam, Nucl. Phys. B377 (1992) 143.}

\Ref\GKLEB{D. Gross and I. Klebanov, Nucl. Phys. B359 (1991) 3.}

\Ref\AJEV{J. Avan and A. Jevicki, Phys. Lett B266 (1991) 35; Phys. Lett.
B272 (1990) 17; Mod. Phys. Lett. A7 (1992) 357.}

\Ref\DDMW{S.R. Das, A. Dhar, G. Mandal and S.R. Wadia, Int. J. Mod. Phys.
A7 (1992); Mod. Phys. Lett. A7 (1992) 71; Mod. Phys. Lett. A7 (1992) 937.}

\Ref\MORSEIB{G. Moore and N. Seiberg, Int. J. Mod. Phys. A7 (1992) 2601.}

\Ref\POLWIN{D. Minic, J. Polchinski and Z. Yang, Nucl. Phys. B369 (1992) 324}

\Ref\DMW{A. Dhar, G. Mandal and S.R. Wadia, TIFR Preprints TIFR-TH/91-61
and TIFR /TH/ 92-40}

\Ref\GKN{D. Gross, I, Klebanov and M. Newmann, Nucl. Phys. B350 (1991)
621; U. Danielsson and D. Gross, Princeton Preprint PUPT-1258 (1991).}

\Ref\POLCO{J. Polchinski, Nucl. Phys. B346 (1990) 253}

\Ref\DIFRKUT{P. Di Francesco and D. Kutasov, Phys. Lett. B261 (1991) 385.}

\Ref\DSTATE{ A. Polyakov, Mod. Phys. Lett. A6 (1991) 635}

\Ref\WGROUND{E. Witten, Nucl. Phys. B373 (1992) 187; I. Klebanov and
A. Polyakov, Mod. Phys. Lett. A6 (1991) 3273}

\Ref\BLACKA{G. Mandal, A. Sengupta and S.R. Wadia, Mod. Phys. Lett. 6
(1991) 1685}

\Ref\BLACKB{E. Witten, Phys. Rev. D44 (1991) 314.}

\Ref\MART{E. Martinec and S. Shatashvili, Nucl. Phys. B368 (1992) 338.}

\Ref\EGUCHI{ T. Eguchi, H. Kanno and S.K. Yang, Newton Institute Preprint
NT-92004 (1992).}

\Ref\DHAR{A. Dhar, G. Mandal and S.R. Wadia, private communication;
A. Dhar, Lectures at Spring Workshop on Superstrings at ICTP
Trieste, April 1992 - to appear in proceedings.}

\Ref\BATEMAN{ {\it Tables of Integral Transforms}, Volume II (Bateman
Manuscript Project) p 145 (McGraw-Hill, 1954).}

\Ref\DVV{R. Dijkgraaf, E. Verlinde and H. Verlinde, Nucl. Phys. B371
(1992) 269. }

\Ref\DMWW{A. Dhar, G. Mandal and S.R. Wadia, (in preparation)}

\Ref\ELL{J. Ellis, N.E. Mavromatos and D. Nanopoulos, Phys. Lett. B272
(1991) 261.}

\Ref\BRUS{J. Ellis, N.E. Mavromatos and D. Nanopoulos, Phys. Lett. B267
(1991) 465; R. Brustein and S. de Alwis, Phys. Lett. B272 (1991) 285.}

In the past two years strings moving in two dimensions have been studied
extensively from two different points of view. The first is a lattice
approach based on the one dimensional matrix model [\DONEG - \GKN],
while the second is
based on the Polyakov formulation of the two dimensional critical string
[\POLCO - \BLACKB].
The double scaling continuum limit of the $d=1$ matrix model with a scaled
inverted harmonic oscillator potential has been shown to correspond to one
background of the two dimensional critical string, viz. the liouville
theory coupled to a single matter field on the worldsheet. Indeed there is
very good agreement of matrix model and continuum results for tree level
quantities like the $S$-matrix. Furthermore both the matrix model and the
one dimensional Liouville theory have the same global $W$ - infinity
symmetry \foot{ In the matrix model the symmetry has been shown to be
present in the full quantum theory and corresponds to $\Win$ [\DDMW]. In
the continuum theory the tree level theory has been shown to possess a
$\win$ symmetry [\WGROUND] which is the classical limit of the
$\Win$ symmtery} [\AJEV - \POLWIN, \WGROUND].

Another background of the continuum
two dimensional critical string has attracted a
lot of attention in recent times for obvious reasons :
that of the two dimensional black hole [\BLACKA, \BLACKB].
However an understanding of the black hole in the matrix model has not yet
been achieved. In this note we provide a hint to the connection of matrix
model with fields moving in a two dimensional black hole background.

A physically transparent description of the perturbative aspects
of the scaling limit of the $d=1$ matrix model,
at least in the low energy sector, is provided by Collective field theory
where the collective field is the density of eigenvalues of the matrix
[\JEVSAK, \DJEV]. In the most direct interpretation of collective field theory,
the
"time-of-flight" variable in the matrix model is identified with the spatial
dimension of the two dimensional critical string around a
linear dilaton and cosmological constant background, and
the fluctuation of the collective field is identified with the space
derivative of the massless tachyon [\DJEV]. This interpretation also
follows directly from the fermionic field theory formulation [\SW].

In this note we show that a certain {\it linear}
integral transform of the fluctuations of the
collective field obeys the same linearized
classical equation of motion as that of a massless scalar
in the black hole background of the two dimensional critical string. By
explicitly evaluating the transform of a general classical solution of the
collective field we show that we get the correct solutions for a massless
scalar in the black hole background in all the regions of spacetime.
Thus the elementary excitation of the matrix model can be also
interpreted as a massless scalar in a black hole background \foot{Our
proposal is different from the speculation in [\ELL] where it was proposed
that a {\it covariantized} collective field theory along the lines of
[\BRUS] could describe tachyons in black hole backgrounds. Here we deal
with the same collective field theory as in [\DJEV] and not an extension
of it.}. The
physically interesting asymptotic states in the black hole background are
related by a nonlocal transformation to those in a linear dilaton -
cosmological constant background so that the physics of scattering
processes are not identical.
Nevertheless quantities of interest in black hole physics may be
calculable from those in matrix model and this connection may provide a
possible framework for understanding non-perturbative issues in the
quantum aspects of black holes.
In the context of worldsheet theories a similar connection
between the $SL(2,R)$ coset model and liouville theory have been
noted earlier in [\MART]. Finally since the transformation is
linear, we argue that
symmetries of string theory in a black hole background are isomorphic to that
in the liouville background, a fact recently found in the continuum theory
[\EGUCHI].

The collective field hamiltonian for the double scaled $d=1$ matrix model
is given by
$$ H = \int dx~ [{1 \over 2 \beta^2} (\pax \Pip (x))~\phi (x)~
(\pax \Pip (x)) + \beta^2 ( {\pi^2 \over 6} \phi^3 (x)
+ (4 \mu - x^2) \phi (x) ) + \Delta V\eqn\one$$
where $x$ denotes the space of eigenvalues of the original matrix model,
$\phi(x)$ denotes the collective field which is the density of eigenvalues
and $\Pip (x)$ is the momentum conjugate to $\phi(x)$. $\mu$ denotes the
scaled fermi level to be identified with the worldsheet cosmological
constant. In \one\ $\beta = {N \over g}$, where the original matrix is $N
\times N$ and the coupling of the matrix model is $g$. In \one\ $\Delta V
= \half \int_{x=y} dx~\phi(x) \partial_x \partial_y {\rm log}~ \vert x - y
\vert$.
In this paper we shall be concerned with the classical theory;
we will, therefore, ignore issues of normal ordering of the above
hamiltonian (which is of course necessary to make contact with the quantum
string theory).

The classical equations of motion follow from the above hamiltonian
$$\pat \phi = - {1 \over \beta^2}\pax[\phi(\pax \Pip)]~~~~~~
\pat \Pi = - {1 \over 2 \beta^2} (\pax \Pip)^2 + \beta^2 ({\pi^2 \over 2}
\phi^2 + (4 \mu - x^2))\eqn\two$$
The perturbative ground state solution is then given by the time
independent configuration
$$ \phi_0 (x) = {\stwo \over \pi} (x^2 - 4\mu)^{\half}~~~~~\Pip (x,t) = 0
\eqn\three$$

We expand around this ground state solution in the standard manner
$$ \phi (x,t) = \phi_0 (x) + {1 \over \beta} \tphi (x,t) ~~~~~
\Pip (x,t) = \beta \tPi (x,t) \eqn\four$$
Consider first the linearized equation satisfied by the fluctuation $\tphi
(x,t)$. This is obtained by eliminating the momenta $\Pip$ from the
equations of motion and reads
$$ \half \pat^2 \tphi = \tphi + 3x \pax \tphi + (x^2 - 4 \mu) \pax^2 \tphi
\eqn\five $$

To obtain the standard interpretation of the collective field one
introduces the time of flight variable $\tau$ and write $\tphi$ in terms
of a new field $\eta(\tau,t)$ [\DJEV]
$$ \tau \equiv \int^x~{dx \over \phi_0}~~~~~~\tphi (x,t) \equiv \pax
\eta (\tau,t) \eqn\seven$$
For our double scaled potential $ x = 2 \smu~cosh~\tau $.
Semi-classically the range of $x$ is from $ 2 \smu$ to $\infty$, so that
the range of $\tau$ is from $0$ to $\infty$. The field $\eta (\tau,t)$
satisfies Dirichlet boundary conditions at $\tau = 0$.
With a trivial rescaling of the time $t$ one has the equation for a
massless scalar field in two dimensions :
$ [\partial_t^2  - \partial_\eta^2] \eta (\tau,t) = 0 $.
As is well known, the interaction
terms reflect the behavior expected from the presence of a nontrivial
dilaton background which asymptotically agrees with a linear dilaton.

For the purposes of this paper it is convenient to work in terms of the
field $\tphi (x,t)$ rather than $\eta (\tau,t)$.
The first step in our transformation is to define
$$ U (p,t) \equiv {1 \over p} \int_{2 \smu}^{\infty} dx~ e^{-px} \tphi (x,t)
\eqn\nine$$
Using the Dirichlet boundary conditions it may be easily checked that the
equation for $\tphi$ implies the following equation for $U(p,t)$
$$ [(p \pap)^2 + 2 p \pap + 1 - \pat^2 - 4 \mu p^2] U(p,t) + O({1 \over
\beta}) = 0 \eqn\ten$$

The quantity $U(p,t)$ is related to the macroscopic loop operator (with
its classical value subtracted out) by a
factor of ${1 \over p}$, so that to this order of $\onbeta$
$p U(p,t)$ satisfies the Wheeler deWitt
equation for the Liouville model. The latter can be also derived directly
from the fermionic formulation of the matrix model \foot{This has been
written down from a calculation of two point functions in [\MORSEIB]. The
loop operator defined with {\it imaginary} $p$ has been shown to obey the
Wheeler-de Witt equation directly from the operator equations of motion of the
fermionic field theory in [\DHAR].}.

We now define a field $T(u,v)$ by the following transform
$$ T(u,v) = \int_0^\infty dp \int_{-\infty}^\infty dt~p~
e^{ip[e^t v + e^{-t} u]} U(p,t) \eqn\eleven$$
The equation $\ten$ for $U(p,t)$ then implies the following equation for
$T(u,v)$
$$ [4(uv + \mu) \pau \pav + 2(u \pau + v \pav) + 1] T(u,v) = 0
\eqn\twelve $$
This is precisely the equation of the massless tachyon moving in a black
hole background of two dimensional critical string theory written in
Kruskal like coordinates. The invariant form of the equation is
$$ \nabla^2 T - 2 \nabla T \cdot \nabla D + (\twa) T = 0 \eqn\thirteen$$
where $\nabla$ denotes target space covariant derivative and $D$ is the
dilaton background. $\alpha '$ is the string tension. In Kruskal like
coordinates the black hole solution in the small $\alpha '$ limit
has the metric and dilaton fields
[\BLACKA]
$$ \eqalign{G_{uv} = & G_{vu} = {1 \over 2(\twa~uv + a)}~~~G_{uu} = G_{vv}
= 0 \cr & D(u,v) = -\half~{\rm log}~(\twa~uv + a)}\eqn\fourteen$$
The parameter $a$ is the mass of the black hole.
Substituting \fourteen\ in \thirteen\ and comparing the resulting equation
with  \twelve\ we get for the black hole mass
$$ a = \twa~\mu \eqn\fifteen$$
The same relation holds in the connection between $SL(2,R)$ coset model
and liouville theory proposed in [\MART].

The field $T(u,v)$ is defined in the entire two dimensional plane. We will
now show explicitly that $T(u,v)$ thus defined gives the correct solution
of \twelve\ in all the regions of the $(u,v)$ plane. To do this it is
convenient to define coordinates $(r, \theta)$ in the four regions as
follows
$$ \eqalign{ & u = r~e^\theta~~~~v = r~e^{-\theta}~~~~{\rm for}~~~u,v
\ge 0~~~~{\rm Region~ I} \cr
& u = -r~e^\theta~~~~v = r~e^{-\theta}~~~~{\rm for}~~~u < 0, v > 0
{}~~~~{\rm Region~II} \cr & u = - r~e^\theta~~~~v = - r~e^{-\theta}
{}~~~~{\rm for}~~~u,v < 0 ~~~~ {\rm Region~III} \cr & u = r~e^\theta
{}~~~~v = - r~e^{-\theta}~~~~{\rm for}~~~u > 0, v < 0
{}~~~~ {\rm Region~IV}} \eqn\fivea$$
Regions I and III are the exterior regions. Region II contains the future
black hole singularity at $uv = a$ while Region IV contains the past
singularity.

Let us now obtain $T(r,\theta)$ by starting from the definition \eleven.
Start with the equation satisfied by $U(p,t)$. A solution of the form
$ U(p,t) = e^{i\nu t} \unu (p) $ is readily obtained. For $\eta (\tau,t)$
satisfying Dirichlet boundary condition at $\tau = 0$ the transform
\eleven\ may be evaluated to yield
$$ \unu (p) = {1 \over p} K_{i\nu} (2 \smu~p) \eqn\fivefive$$
where $K$ stands for the modified Bessel function. Using the
parametrization in equation \fivea\ and performing the integration over
$t$ (after a shift of the integration variable $t \rightarrow t + \theta$
we find
$$ \eqalign{& T(r,\theta) = 2 e^{-i\nu \theta} \int_0^\infty dp~K_{i\nu}(2
\smu p)~K_{i\nu} (-2ipr)~~~~~~~\rega \cr &
T(r,\theta) = 2 e^{-i\nu \theta} \int_0^\infty dp~K_{i\nu}(2
\smu p)~K_{i\nu} (2pr)~~~~~~\regb}\eqn\fiveseven$$
We will not write down formulae for the other two regions since they are
trivially related to those in I and II.
The $T(r,\theta)$ are thus given in terms of $K$-transforms of the
macroscopic loop operator. Both the above integrals may be evaluated
explicitly [\BATEMAN]. The final result is
$$\eqalign{ &T(r,\theta) = {\pi^2~e^{-{\pi \nu \over 2}} \over
4~cosh~\pi\nu} ~\mu^{{i\nu \over 2}}~e^{-i\nu \theta}~(r)^{-(1 +
i\nu)}~F(\half + i\nu, \half; 1; 1 + {\mu \over r^2})~~~~~\rega \cr &
T(r,\theta) = {\pi^2~e^{-{\pi \nu \over 2}} \over 4~cosh~\pi\nu}
{}~\mu^{{i\nu \over 2}}~e^{-i\nu \theta}~(r)^{-(1 +
i\nu)}~F(\half + i\nu, \half; 1; 1 - {\mu \over
r^2})~~~~~\regb}\eqn\fiveeight$$
These are in exact agreeement with one solution of \twelve\ in each of the
regions. This may be seen by writing
$$ T(r,\theta) = e^{i \omega \theta}~r^{-(1 + i\omega)}
{}~T_{\omega}(r) \eqn\fiveb$$
Defining new variables $z = 1 + {\mu \over r^2}$ in
Regions I and III and $z = 1 - {\mu \over r^2}$ in Regions II and IV, we
obtain the a standard hypergeometric equation. One solution may be written
as
$$\eqalign{&\tom (r) = F(\half + i \omega, \half; 1; 1 + {\mu \over r^2})
{}~~~~~~\rega \cr &
\tom (r) = F(\half + i \omega, \half; 1; 1 - {\mu \over r^2})
{}~~~~~~\regb} \eqn\fivefour$$
where $F(a,b;c;z)$ stands for the hypergeometric function.
The solutions in Regions III and IV are obtainable
as trivial extensions of the solutions in Regions I and II. The other
solution for $\tom (r)$ may be obtained in the standard manner
$$\eqalign{\tom '(r) = &\Gamma (\half +i\omega)~ \Gamma (\half)~
F(\half + i\omega,\half;1; 1 + {\mu \over r^2})~ {\rm log}~(1 + {\mu \over
r^2}) + \cr & \sum_{m=1}^{\infty} s_m {\Gamma (\half + m + i\omega)
\Gamma (\half + m) \over (m !)^2}~(1 + {\mu \over r^2})^m
{}~~~~\rega }\eqn\sixone$$
where $s_m = \sum_{n=1}^m {1 \over n + i\omega - \half} +
{1 \over n - \half} - {2 \over n}$.
In Region II one has to replace $\mu \rightarrow -\mu$. The solutions
\fivefour\ agree with the expressions \fiveeight\ (upto a constant).

It is interesting that the $T(r,\theta)$ obtained by evaluating the
transform yields {\underbar{one}}
of the solutions of the differential equation
satisfied by $T(u,v)$. This is related to the fact that the matrix model
always picks out a {\it specific} combination of dressings of the
continuum theory, as is also manifest in the standard interpretation of
the matrix model as a liouville background. Perhaps more significantly the
solution in the interior region which is picked out is the one which is
regular at the singularity $r^2 = \mu$ in Region II, and not the one which
has a logarithmic singularity \foot{The behavior of the tachyon field near
the singularity has been investigated in [\BLACKA] and [\DVV].}.

To obtain the asymptotic states which satisfies physically interesting
boundary conditions in the exterior region it is necessary to rewrite
the above solutions using standard relations between hypergeometric
functions. For example in Region I
$$\eqalign{T(r,\theta) = &{e^{-i\nu \theta} \over r}
[({\mu \over r^2})^{i \nu \over 2}
A(-\nu) F( \half + i\nu, \half; 1 + i\nu; -{\mu \over r^2}) + \cr &
(-1)^{i\nu} ({\mu \over r^2})^{-i \nu \over 2} A(\nu) F(
\half - i\nu, \half; 1 - i\nu; -{\mu \over r^2})]} \eqn\fivenine$$
where $A(\nu) \equiv {\Gamma(i\nu) \over \Gamma(\half + i \nu) \Gamma
(\half)}$ and we have omitted an overall $\nu$-dependent constant. The two
terms in \fivenine\ are in exact agreement with two of the
tachyon vertex operators found in the $SL(2,R)/U(1)$ coset model in [\DVV]
after a change of variables $r \rightarrow {\rm sinh}~(r/2)$. These modes
vanish on the past and future null infinities respectively and represent
left and right moving modes at spatial infinity. The other two modes which
vanish at the past and future horizons ( and represent left and
rightmoving plane waves on the horizon) are given by the two terms of the
following rewriting of the hypergeometric functions in \fiveeight\
$$\eqalign{T(r,\theta) = &{e^{-i\nu \theta} \over r}
[({\mu \over r^2})^{i \nu \over 2}
A(\nu) F( \half - i\nu, \half; 1 - i\nu; -{r^2 \over \mu}) + \cr &
(-1)^{i\nu} ({\mu \over r^2})^{-i \nu \over 2} A(-\nu) F(
\half + i\nu, \half; 1 + i\nu; -{r^2 \over \mu})]} \eqn\fiveten$$
Thus the transform of the macroscopic loop operator is a linear
combination of the asymptotic solutions in the black hole geometry. This
fact should play an important role in understanding quantum aspects of the
black hole in terms of matrix model quantities.

It is convenient to view $T(u,v)$ as a fourier transform of a different
field, particularly if one is interested in performing the transformation
in the action. This is defined as follows. Consider coordinates $(z,w)$
which are related to $(p,t)$ in the following manner:
$$ \eqalign{ & z = \pm p~e^t~~~~w = \pm p~e^{-t}~~~~{\rm for}~~~z,w \ge 0~~
({\rm Region~ A})~ {\rm and}~ z,w < 0 ~({\rm Region~ C})  \cr
& z = \mp p~e^t~~~~w = \pm p~e^{-t}~~~~{\rm for}~~~z < 0, w > 0 ~~
({\rm Region~B})~{\rm and}~ z > 0, w < 0~~({\rm Region~ D})} \eqn\fourta$$
The coordinates $(z,w)$ thus cover the entire two dimensional plane.
Then we define a field ${\bar U} (z,w)$ as
$$ \eqalign{& {\bar U} (z,w) = U(p,t) ~~~~~~~~{\rm in~ Region~A~and~C} \cr
& {\bar U} (z,w) = U(ip, t) ~~~~~~~~{\rm in~ Region~B~and~D}}\eqn\fourtb$$
Note that $U(ip,t)$ satisfies equation \ten\ with $\mu \rightarrow - \mu$.
Then the transformation \eleven\ is a simple fourier transform
of ${\bar U} (z,w)$ thus defined in the
the entire $(z,w)$ plane to the $(u,v)$ plane. One may now use this to
write down the quadratic action of $U(z,w)$ starting from the quadratic
action for $T(u,v)$ or vice versa. Note that due to differing signs in the
jacobians for transformations from $(z,w)$ to $(p,t)$ in the different
regions the contributions of the various regions actually {\it add} up
when expressed as an integral over $p$ and $t$.

It is now well known that the $d=1$ matrix model has
a $\Win$ symmetry, which is most transparent in the fermionic formulation
of the model [\DDMW, \MORSEIB].
In the classical collective field theory the corresponding
symmetry is the classical limit  of $\Win$ symmetry, viz. the $\win$
symmetry [\AJEV, \POLWIN].
It is as yet unknown whether this $\win$ symmetry becomes the
full quantum $\Win$ symmetry in the collective field theory \foot{In
[\AJEV] it was argued that the symmetry of the quatum collective field
theory is still $\win$. This is in contradiction with the exact results of
the matrix model, as obtained from the fermionic field theory [\DDMW].}.
Our identification of the tachyon field around a black hole as a {\it linear}
integral transform of the collective field shows that the classical theory
around a black hole background would also possess a $\win$ symmetry.
Indeed recent studies of the $SL(2,R) / U(1)$ conformal field theory seem
to indicate an isomorphism of the symmetry structure with that of the
$c=1$ liouville theory [\EGUCHI].
Our result provides a natural explanation of this
isomorphism from the point of view of the matrix model.

To enable one to use the full power of the matrix model to address
questions in black hole physics it is necessary to express the above
transformation in the language of the fermionic field theory. The basic
object in the fermionic formulation is the bilocal operator of fermions,
one version of which is [\DDMW]
$$W(\alpha,\beta,t)= 1/2 \int_{-\infty}^\infty dx\; e^{i\alpha x}\;
\psi^\dagger(x+\beta/2,t)\; \psi(x-\beta/2,t) \eqn\sixteen$$
Clearly $W(\alpha,0,t)$ is the fourier transform of the collective field
$\phi (x,t) = \psi^\dagger (x,t) \psi (x,t)$ and is related to the
quantity $p U(p,t)$ for imaginary values of $p$. The equal time
commutation relations of the $W(\alpha, \beta,
t)$ satisfy the full $\Win$ algebra. In the classical limit this reduces
to the $\win$ algebra, the algebra of area preserving diffeomorphisms on a
plane.

In view of the above discussion it is natural to define the quantities
$$ Z (u, \beta, v) \equiv \int_{-\infty}^{\infty}
d\alpha~dt~~e^{i\alpha[e^t~v + e^{-t}~u]}
W(\alpha, \beta,t) \eqn\seventeen $$
which would obey an isomorphic algebra to that of $W(\alpha,\beta,t)$. The
quantities $Z(u, \beta, v)$ would then represent the natural generators of
the algebra in the new language.

The complete bosonization of the fermionic field theory is best expressed
in terms of the quantum distribution function $u(x,p,t)$ which is the
fourier transform of the $W(\alpha,\beta,t)$ [\DMW]
$$ u(x,p,t) = \int d\alpha d \beta~e^{i\alpha x + i\beta p}
W(\alpha,\beta,t) \eqn\eighteen$$
The integral of $u(x,p,t)$ over $p$, $\int dp~u(x,p,t)$ being simply the
collective field $\phi(x,t)$.
In [\DMW] the matrix model is exactly rewritten in terms of the field
$u(x,p,t)$ ; this bosonic model reduces to the collective field theory in
an appropriate limit.
Our discussion then indicates that the natural bosonic
variable in terms of which
the matrix model appears to be a critical string theory in a black hole
background is
$$ s(u,p,v) = \int dl~dt~e^{il[e^t v + e^{-t} u]} \int^{\infty}_{2\smu} dx
{}~e^{-lx}~u(x,p,t) \eqn\nineteen$$
Perhaps a genuinely three dimensional transform from the $(x,p,t)$ is more
appropriate.

In conclusion, we have shown that the
the $d=1$ matrix model can be interpreted as a theory of the
massless tachyon moving in the black hole background of the two
dimensional critical string theory, the tachyon field being a {\it linear}
integral transform of the collective field.
While we have shown this from the collective field theory point of view
the connection should be understandable in the fermionic field theory as
well, the latter being more suitable for nonperturbative studies.
The important question is to understand the exact calculations of the
matrix model in the framework of this interpretation.
The connection we found may lead to the
exciting possibility that vexing questions in
quantum black holes may be addressed and resolved in an unambiguous way by
posing the right questions in the matrix model.

\centerline{ACKNOWLEDGEMENTS}

I would like to thank A. Dhar,
G. Mandal, A. Sen and A. Sengupta for discussions.
I am grateful to S. Wadia for infecting me with his conviction that black
holes should be describable in matrix model and for useful discussions.
I would also like to thank A. Dhar, A. Sen and S. Wadia for a critical
reading of the manuscript.

\noindent{Note Added:~~~~~~} An understanding of the connection between
matrix model and black hole in the framework of the bosonized theory of
Ref. [\DMW] has been achieved [\DMWW].

After submitting this paper, E. Martinec has pointed out that the
published version of [\MART] has an added note in which it is shown that
the solutions to the tachyon field in a black hole background can be
obtained from  solutions to the Wheeler-de Witt equation by the integral
transform discussed in this paper. In our treatment we start with the
collective field of the matrix model and obtain one {\it specific}
combination of the solutions. I wish to thank E. Matrinec for pointing
this out to me.

\refout
\end